# A New Generic Taxonomy on Hybrid Malware Detection Technique

Robiah Y, Siti Rahayu S., Mohd Zaki M, Shahrin S., Faizal M. A., Marliza R.
Faculty of Information Technology and Communication
Univeristi Teknikal Malaysia Melaka,
Durian Tunggal, Melaka,
Malaysia
robiah@utem.edu.my, sitirahayu@utem.edu.my, zaki.masud@utem.edu.my, shahrinsahib@utem.edu.my,
faizalabdollah@utem.edu.my, marliza@utem.edu.my

*Abstract-Malware is a type of malicious program that replicate from host machine and propagate through network. It has been considered as one type of computer attack and intrusion that can do a variety of malicious activity on a computer. This paper addresses the current trend of malware detection techniques and identifies the significant criteria in each technique to improve malware detection in Intrusion Detection System (IDS). Several existing techniques are analyzing from 48 various researches and the capability criteria of malware detection technique have been reviewed. From the analysis, a new generic taxonomy of malware detection technique have been proposed named Hybrid-Malware Detection Technique (Hybrid-MDT) which consists of Hybrid-Signature and Anomaly detection technique and Hybrid-Specification based and Anomaly detection technique to complement the weaknesses of the existing malware detection technique in detecting known and unknown attack as well as reducing false alert before and during the intrusion occur.*

## I  INTRODUCTION

Malware is considered as worldwide epidemic due to the malware author's activity to have a finance gain through theft of personal information such as gaining access to financial accounts. This statement has been proved by the increasing number of computer security incidents related to vulnerabilities from 171 in 1995 to 7,236 in 2007 as reported by Computer Emergency Response Team [1]. One of the issues related to this vulnerability report is malware attack which has generated significant worldwide epidemic to network security environment and bad impact involving financial loss.

Hence, the wide deployment of IDSs to capture this kind of activity can process large amount of traffic which can generate a huge amount of data. This huge amount of data can exhaust the network administrator's time and implicate cost to find the intruder if new attack outbreak happen especially involving malware attack. An important problem in the field of intrusion detection is the management of alerts as IDS tends to produce high number of false positive alerts [2]. In order to increase the detection rate, the use of multiple IDSs can be used and correlate the alert but in return it increases the number of alerts to process. Therefore certain detection mechanisms or technique need to be integrated with IDS correlation process in order to guarantee the malware is detected in the alert log. Hence, the proposed research is to generate a new generic taxonomy of malware detection technique that will be the basis of developing new rule set for IDS in detecting malware to reduce the number of false alarm.

The rest of the paper is structured as follows. Section II discuses the related work on malware and the current taxonomy of malware detection technique. Sections III present the classification and the capability criteria of malware detection techniques. Section IV discusses the new propose taxonomy of malware detection technique and. Finally, section V conclude and summarize future directions of this work.

## II  RELATED WORK

### A.  What is Malware?

According to [3], malware is a program that has malicious intention. Whereas [4] has defined it as a generic term that encompasses viruses, Trojans, spywares and other intrusive codes. Malware is not a "bug" or a defect in a legitimate software program, even if it has destructive consequences. The malware implies malice of forethought by malware inventor and its intention is to disrupt or damage a system.

[5] has done research on malware taxonomy according to their malware properties such as mutually exclusive categories, exhaustive categories and unambiguous categories. In his research he has stated that generally malware is consists of three types of malware of the same level as depicted in Figure 1 which are virus, worm and Trojan horse although he has commented that in several cases these three types of malware are defined as not being mutually exclusive

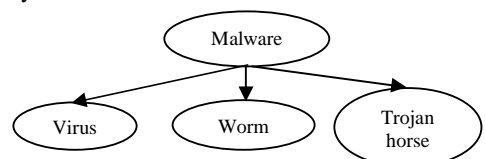

Figure 1. General Malware Taxonomy by Karresand

### B. What is Malware Intrusion Detector?

Malware intrusion detector is a system or tool that attempts to identify malware [3] and contains malware before it can reach a system or network. Diverse research has been done to detect this malware from spreading on host and network. These detectors will use various combinations of technique, approach and method to enable them to detect the malware effectively and efficiently during program execution or static. Malware intrusion detector is considered as one of the component of IDS, therefore malware intrusion detector is a complement of IDS.

### C. What is Taxonomy of Malware Detection Technique?

To clearly identify the malware detection technique terms in depth, a research on a structured categorization which is call as taxonomy is required in order to develop a good detection tools. Taxonomy is defined in [6] as "a system for naming and organizing things, especially plants and animals, into groups which share similar qualities".

[7] has done a massive survey on malware detection techniques done by various researchers and they have come out with taxonomy on classification of malware detection techniques which have only two main detection technique which are signature-based detection and anomaly-based detection. They have considered the specification-based detection as sub-family of anomaly-based detection. The researcher has done further literature review on 48 various researches on malware detection technique to verify the relevancies of the detection technique especially the hybrid malware detection technique so that it can be mapped into the proposed new generic taxonomy of malware detection technique. Refer to Table IV for the mappings of the literature review with the malware detection technique.

### III  CLASSIFICATION OF MALWARE DETECTION TECHNIQUES

Malware detection technique is the technique used to detect or identify the malware intrusion. Generally, malware detection technique can be categorized into Signature-based detection, Anomaly-based detection and Specification-based detection.

### A. Overview of Detection Technique

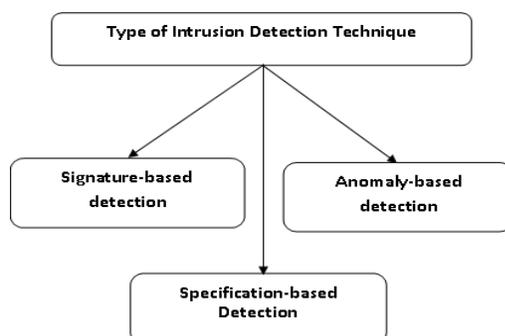

Figure 2. Existing taxonomy of malware detection technique

According to [8] and [9], intrusion detection technique can be divided into three types as in Figure 2 which are signature-based or misuse detection, anomaly-based detection and specification-based detection which shall be a major reference in these research. Based on previous worked [8][9][10][11], the characteristics of each techniques are as follows.

### B. Signature-based detection

Signature-based or sometime called as misuse detection as described by [10] will maintain database of known intrusion technique (attack signature) and detects intrusion by comparing behavior against the database. It shall require less amount of system resource to detect intrusion. [8] also claimed that this technique can detect known attack accurately. However the disadvantage of this technique is ineffective against previously unseen attacks and hence it cannot detect new and unknown intrusion methods as no signatures are available for such attacks.

### C. Anomaly-based detection

Anomaly-based detection stated by [10] analyses user behavior and the statistics of a process in normal situation, and it checks whether the system is being used in a different manner. In addition [8] has described that this technique can overcome misuse detection problem by focusing on normal system behavior rather than attack behavior. However [9] assume that attacks will result in behavior different from that normally observed in a system and an attack can be detected by comparing the current behavior with pre-established normal behavior.

This detection approach is characterized by two phases which is the training phase and detection phase. In training phase, the behavior of the system is observed in the absence of attack, and machine learning technique is used to create a profile of such normal behavior. In detection phase, this profile is compared against the current behavior, and deviations are flagged as potential attacks. The effectiveness of this technique is affected by what aspect or a feature of system behavior is learnt and the hardest challenge is to be able to select the appropriate set of features.

The advantage of this detection technique is that it can detect new intrusion method and capable to detect novel attacks. However, the disadvantage is that it needs to update the data (profiles) describing the user's behavior and the statistics in normal usage and therefore it tend to be large and therefore need more resources, like CPU time, memory and disk space. Moreover, the malware detector system often exhibit legitimate but previously unseen behavior, which leads to high rate of false alarm.

### D. Specification-based detection

Specification-based detection according to [9] will rely on program specifications that describe the intended behavior of security-critical programs. The goal of the policy specification language according to [11] is to provide a simple way on specifying the policies of privileged programs.

It monitors executions program involve and detecting deviation of their behavior from the specification, rather than detecting the occurrence of specific attack patterns. This technique is similar to anomaly detection where they detect the attacks as deviate from normal.

The difference is that instead of relying on machine learning techniques, it will be based on manually developed specifications that capture legitimate system behavior. It can be used to monitor network components or network services that are relevant to security, Domain Name Service, Network File Sharing and routers.

The advantage of this technique according to [8] is that the attacks can be detected even though they may not previously encounter and it produce low rate of false alarm. They avoid high rate of false alarm caused by legitimate-but-unseen-behavior in anomaly detection technique. However, the disadvantage is that it is not as effective as anomaly detection in detecting novel attacks, especially involving network probing and denial-of-service attacks due to the development of detail specification is time-consuming and hence increase false negative due to attacks may be missed. Table I summarized the advantages and disadvantages of each technique.

<div align="center">TABLE I</div>
<div align="center">Comparison of Malware detection techniques</div>

| Technique | Advantage | Disadvantage |
|---|---|---|
| Signature-based | • Can detect known attack accurately.<br>• Less amount of system resource is required to detect intrusion.<br>• Focus on attack behaviour | • It cannot detect new, unknown intrusion methods.<br>• Ineffective against previously unseen attacks, as no signatures are available for such attacks. |
| Anomaly-based | • It can detect new intrusion method and novel attack.<br>• Focus on normal behaviour to overcome undetected unknown attack. | • It needs to update the data describing the users behaviour and the statistics in normal usage (profiles) and it tend to be large.<br>• Problem to select the appropriate set of features to be able to detect potential attacks.<br>• Need more resources, like cpu time, memory and disk space.<br>• High false positive alarm. |
| Specification-based | • Attacks can be detected even though they may not previously encounter<br>• It has low rate of false positive alarm. | • It is not as effective as anomaly detection in detecting novel attacks; especially in network probing and denial-of-service attacks.<br>• Development of detail specification is time-consuming.<br>• Increase false negative due to attacks may be missed. |

*E. Proposed criteria for Malware Detection Technique*

Three major detection techniques have been reviewed and the objective of this research is to develop a new generic taxonomy on malware detection technique. It can be done by analyzing the current malware detection technique and identify the significant criteria within each technique that can improve the IDS problem. As mentioned by [12], IDS has developed issues on alert flooding, contextual problem, false alert and scalability. The characteristic that shall be analyzed in each detection technique is according to the issue listed in Table II.

<div align="center">TABLE II</div>
<div align="center">Issue analyzed in IDS</div>

| No | IDS Issue | Description | Propose Solution |
|---|---|---|---|
| 1 | Alert flooding | IDS are prone to alert flooding as they provide a large number of alerts to the network security officer, who then has the difficulties coping with the load. | • To reduce number of alert generated from IDS. |
| 2 | Contextual problem | Attacks are likely to generate multiple related alerts. Current IDS do not make it easy for network security officers to logically group related alerts. | • To identify multi-step attack |
| 3 | False Alert | Existing IDS are likely to generate false positives or false negatives alerts | • To reduce number of false positive alerts<br>• To reduce number of false negative alerts<br>• To identify known attack using misuse detection<br>• To identify unknown attack using anomaly detection |

[13] has proposed the criterion of malware detection technique that shall be analyzed against the issue listed in Table II , which are :-

1. Capability to do alert reduction
2. Capability to identify multi-step attack.
3. Capability to reduce false negative alert.
4. Capability to reduce false positive alert
5. Capability to detect known attack
6. Capability to detect unknown attack

Alert reduction is required in order to overcome the problem of alert flooding or large amount of alert data generated by the IDS. This capability criterion is important in order to reduce the network security officer's tension in performing troubleshooting when analyzing the exact attacker in their environment.

For second criteria, most of the malware detection technique is incapable to detect multi-step attack. Therefore this capability is required as attacker behavior is becoming more sophisticated and it shall involve one to many, many to one and many to many attacks.

The third and fourth criteria, most of the IDS have the tendency to produce false positive and false negative alarm.

This false alarm reduction criterion is important as it closely related to alert flooding issue. For fifth and sixth criterion, the capability to detect both known and unknown attack is required to ensure that the alert generated will overcome the issue of alert flooding and false alert.

## IV    DISCUSSION AND ANALYSIS OF MALWARE DETECTION TECHNIQUES

In the current trend, few researches such as [14], [15], [16], [17] and [8] have been found to manipulate this detection technique by combining either Signature-based with Anomaly-based detection technique(Hybrid-SA) or Anomaly-based with Specification-based detection technique (Hybrid-SPA) in order to develop an effective malware detector's tool.

In this paper, a new proposes taxonomy of malware detection technique is proven to be effective by matching the current malware detection technique: Signature-based detection, Anomaly-based detection and Specification-based detection with capability criteria propose by [13] as discussed in section III. This analysis is summarized in Table III.

TABLE III
Malware detection technique versus proposed capability criteria
(Capable=√, incapable=×)

| No | Technique Name | Alert Reduction | Multi-step Attack | Reduce False negative Alert | Reduce False positive Alert | Detect Known Attack | Detect Unknown Attack |
|----|----------------|-----------------|-------------------|------------------------------|------------------------------|---------------------|------------------------|
| 1 | Signature-based | × | × | × | √ | √ | × |
| 2 | Anomaly-based | × | √ | √ | × | √ | √ |
| 3 | Specification-based | × | × | × | √ | √ | √ |

Referring to Table III, all of the detection techniques have the same capability to detect known attack. However, anomaly-based and specification-based have the additional capabilities to detect unknown attack. Anomaly-based has the extra capabilities compare to other detection techniques in terms of reducing false negative alert and detecting multi-step attack. Nevertheless, it cannot reduce the false positive alert which can only be reduced by using signature-based and specification-based technique.

Due to the incapability to reduce either false negative or false positive alert, all of these techniques are incapable to reduce false alert. This has given an implication that there are still some rooms for improvement in reducing false alarm. Based on the analyses, the researcher has propose an improved solution for malware detection technique which can either use combination of signature-based with anomaly-based detection technique (Hybrid-SA) or specification-based with anomaly-based detection technique (Hybrid-SPA) to complement each other weaknesses.

These new technique is later on named by the researcher as Hybrid-Malware Detection Technique (Hybrid-MDT) which shall consists of Hybrid-SA detection and Hybrid-SPA detection technique as depicted in Figure 3.

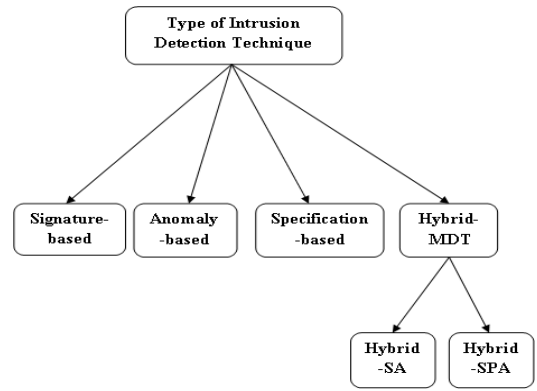

Figure 3. Proposed generic taxonomy of malware detection technique

To further verify the relevancies of the above proposed generic taxonomy of malware detection technique, the researchers have review on 48 researches of various malware detection techniques which can be mapped to the propose taxonomy in Figure 3. Table IV shows the related literature review in malware detection techniques.

TABLE IV
Related literature review in malware detection techniques

| Detection Technique | Paper Review Reference No. |
|---------------------|----------------------------|
| Signature-based | [3], [18], [25], [28], [30], [33], [35], [39], [43], [46], [47], [50] |
| Anomaly-based | [9], [10], [11], [21], [22], [23], [26], [28], [31], [53], [54] |
| Specification-based | [4], [14], [15], [20], [24], [27], [32], [34], [36], [37], [38], [40], [41], [42], [44], [45], [48], [49], [51], [52], [55] |
| Hybrid-SA | [8], [16], [17] |
| Hybrid-SPA | [19] |

## V    CONCLUSION AND FUTURE WORKS

In this study, the researchers have reviewed and analyzed the existing malware detection techniques and match it with the capability criteria propose by [13] to improve the IDS's problem. From the analysis researcher has proposed a new generic taxonomy of malware detection techniques which is called Hybrid-Malware Detection Technique (Hybrid-MDT) which consists of Hybrid-SA detection and Hybrid-SPA detection technique. Both techniques in Hybrid-MDT shall complement the weaknesses found in Signature-based, Anomaly-based and Specification-based technique. This research is a preliminary worked for malware detection. This will contribute ideas in malware detection technique field by generating an optimize rule set in IDS. Hence, the false alarm in the existing IDS will be reduced.